%% file: main.tex
\documentclass[amssymb,prb,twocolumn,superscriptaddress,floats,showkeys,showpacs]{revtex4-2}
\usepackage[T1]{fontenc}
\usepackage{bm}
\usepackage{graphicx}
\usepackage{amssymb}
\usepackage{leftidx}
\usepackage{upgreek}
\usepackage{siunitx}
\usepackage{amsfonts}
\usepackage{amsmath} 
\usepackage{hyperref} 
\usepackage{multirow}
\usepackage{makecell}
\usepackage{xcolor}
\usepackage{xspace}
\hypersetup{colorlinks,citecolor=blue, filecolor=blue ,linkcolor=blue , urlcolor=blue, pdftex}

\usepackage{tabularx}

\begin{document}

\newcommand{\HfS}{$\text{HfS}_{2}$\xspace}

\title{The effect of temperature and excitation energy on Raman scattering in bulk \HfS}

% {author & affiliation}

\def \FUW{Institute of Experimental Physics, Faculty of Physics, University
of Warsaw, Pasteura 5, 02-093 Warsaw, Poland}
\def \China{Hefei Innovation Research Institute, School of Microelectronics, Beihang University, Hefei 230013, P. R. China}
\def \CENT{Centre of New Technologies, University of Warsaw, Banacha 2c, 02-097 Warsaw, Poland}
\def \Wroclaw{Department of Semiconductor Materials Engineering, Faculty of Fundamental Problems of Technology, Wrocław University of Science and Technology, Wybrzeże Wyspiańskiego 27, 50-370, Wrocław, Poland}
\def \Spain{Geosciences Barcelona (GEO3BCN), CSIC, Lluís Solé i Sabarís s.n., 08028, Barcelona, Catalonia, Spain}
\def \MagdaG{Department of Materials Science and Engineering, National University of Singapore, 117575, Singapore} 
\def \MagdaGk{Institute for Functional Intelligent Materials, National University of Singapore, 117544, Singapore}

\author{Igor Antoniazzi} 
\email{igor.antoniazzi@fuw.edu.pl}
\affiliation{\FUW}
\author{Natalia Zawadzka}
\affiliation{\FUW}
\author{Magdalena Grzeszczyk}
\affiliation{\FUW}
\affiliation{\MagdaGk}
\author{Tomasz Woźniak}
\affiliation{\Wroclaw}
\author{Jordi~Ib{\'a}{\~n}ez}
\affiliation{\Spain}
\author{Zahir Muhammad}
\affiliation{\China}
\author{Weisheng Zhao}
\affiliation{\China}
\author{Maciej R. Molas}
\affiliation{\FUW}
\author{Adam Babiński}\email{adam.babinski@fuw.edu.pl}
\affiliation{\FUW}

\begin{abstract} 
Raman scattering (RS) in bulk hafnium disulfide (\HfS) is investigated as a function of temperature (5~K $-$ 350~K) with polarization resolution and excitation of several laser energies. 
An unexpected temperature dependence of the energies of the main Raman-active (A$_{\textrm{1g}}$ and E$_{\textrm{g}}$) modes with the temperature-induced blueshift in the low-temperature limit is observed.
The low-temperature quenching of a mode $\omega_1$ (134 cm$^{-1}$) and the emergence of a new mode at approx. 184 cm$^{-1}$, labeled Z, is reported.
The optical anisotropy of the RS in \HfS is also reported, which is highly susceptible to the excitation energy.
The apparent quenching of the A$_{\textrm{1g}}$ mode at $T$=5~K and of the E$_{\textrm{g}}$ mode at $T$=300~K in the RS spectrum excited with 3.06~eV excitation is also observed.
We discuss the results in the context of possible resonant character of light-phonon interactions.
Analyzed is also a possible effect of the iodine molecules intercalated in the van der Waals gaps between neighboring \HfS layers, which inevitably result from the growth procedure.

\end{abstract}
%\keywords{\HfS, temperature, excitation energy, Raman scattering, anisotropy, transition metal dichalcogenides}

\maketitle

%%%%%%%%%%%%%%%%%%%%%%% {INTRO} %%%%%%%%%%%%%%%%%%%%%%%%%%%%%%%%%
\section{Introduction \label{sec:Intro}}
Layered transition metal dichalcogenides (TMD) and, in particular, their few-layer structures have been drawing the attention of researchers for more than a decade now. 
Although about 60 layered TMDs have been recognized, until now the attention of researchers has focused mainly on Mo- and W-based compounds. 
In spite of the fact that the basic properties of several TMDs in their bulk form are known, \cite{wilson1969} there are many research questions to be addressed.
Hafnium disuplphide (\HfS) is a layered van der Waals (vdW) TMD, which is a member of group IVB.~\cite{cingolani1987raman, roubi1988resonance}
\HfS is a semiconductor characterized by an indirect band gap with an energy of about 2~eV.~\cite{GREENAWAY1965, TERASHIMA1987, Gaiser2004, Zelewski2017}\
Recently, \HfS has been shown to have a very effective electrical response \cite{kanazawa2016transistor} being quoted to the development of thermoelectric and optoelectronic devices.\cite{singh2019}
However, the \HfS structure is highly susceptible to temperature changes\cite{peng2021} and pressure changes,\cite{ibanez2018high, Grzeszczyk2022} which justify the need to uncover the basic properties of this material.

In this work, we address the effect of temperature and excitation energy on Raman scattering (RS) in bulk \HfS. 
Polarization-sensitive RS measurements are also reported.
We observe an unexpected temperature dependence of the energies of main Raman-active (A$_{\textrm{1g}}$ and E$_{\textrm{g}}$) modes with the temperature-induced blueshift in the low-temperature limit.
We report on a specific result of the high-energy 3.06~eV excitation on the RS spectral lineshape with the apparent quenching of the A$_{\textrm{1g}}$ mode at $T=5K$ and of the E$_{\textrm{g}}$ mode at $T=300K$.  
The polarization-dependent RS measurements show that the mode Z observed in the RS at low temperature appears in the spectrum only for the cross-linear configuration. 
We discuss the results and we propose an explanation of the observed data with the aid of first principles calculations.

%%%%%%%%%%%%%%%%%% {Results} %%%%%%%%%%%%%%%%%
\section{Experimental results\label{sec:Results}}

%%%%%%%%%%%%{Room temperature and phonon dispersion}%%%%%%%%%%%
\subsection{Raman scattering in \HfS at room temperature}

\begin{figure}[t]
    \centering
    \includegraphics[width=1\linewidth]{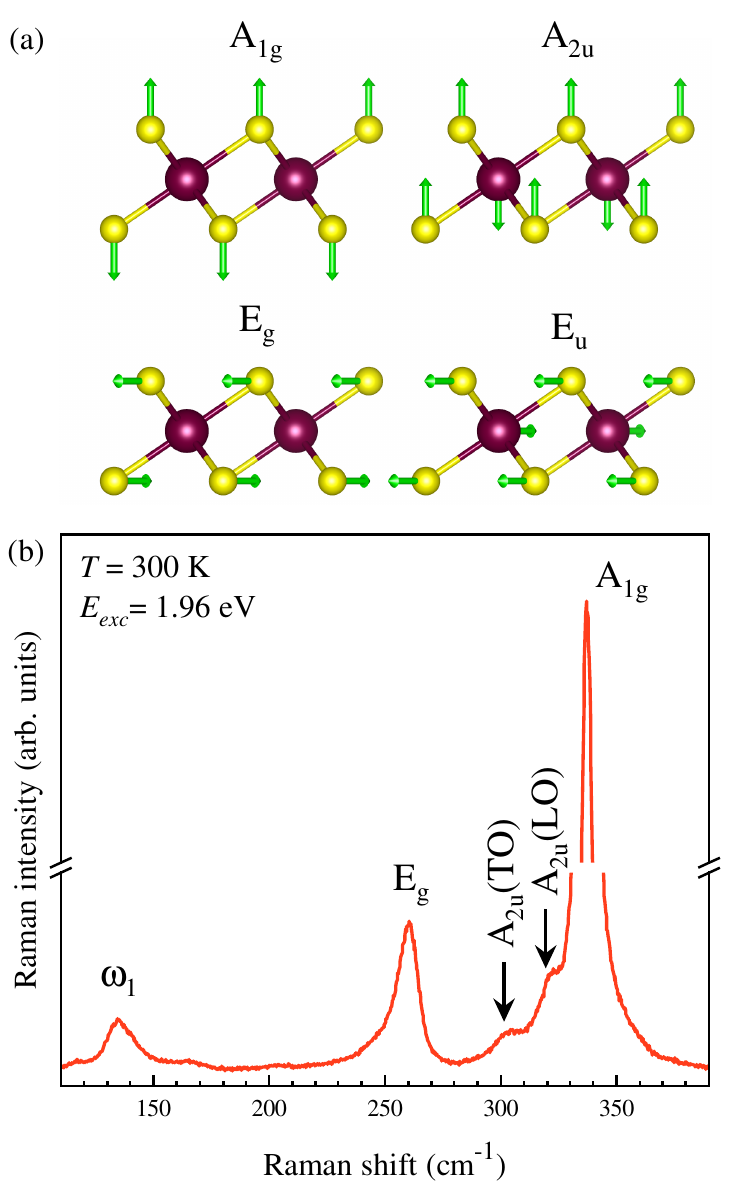}
    \caption
    {\label{fig:Raman_300K}
    (a) Schematic representation of the Raman- (A$_\textrm{1g}$ and E$_\textrm{g}$) and IR-active (A$_\textrm{2u}$ and E$_\textrm{u}$) vibrational modes in 1T phase of \HfS.  
    Yellow/purple balls represent S/Hf atoms.
    Atom displacements of the modes is represented by green arrows. 
    (b) Representative, upolarized RS spectrum of bulk \HfS measured at room temperature under $E_{exc}$= 1.96 eV ($\lambda$= 633 nm).}
\end{figure}

\HfS in its 1T form belongs to the space group $P\overline{3}m1[164]$ with the point group $D_{3d}$. 
There are six normal vibrational modes at the $\Gamma$ point of 1T-\HfS \cite{lukovsky1973IR}: \\
\vspace{-35pt}
\begin{center}
\begin{equation*}
\Gamma=\textrm{A}_\textrm{1g} + \textrm{E}_\textrm{g} + 2\textrm{A}_\textrm{2u} + 2\textrm{E}_\textrm{u}
\end{equation*}
\end{center}

\noindent Schematic representations of the vibrational modes in \HfS are presented in Fig.~\ref{fig:Raman_300K}(a).
However, only two modes, A$_\textrm{1g}$ and E$_\textrm{g}$, are Raman-active.
The infrared-active (IR-active) modes A$_\textrm{2u}$ and E$_\textrm{u}$, split into LO and TO branches.

The representative, unpolarized RS spectrum of the \HfS bulk excited with $E_{exc}$=1.96~eV at 300~K is shown in Fig.~\ref{fig:Raman_300K}(b).
There are five apparent RS peaks in the spectrum. 
According to the literature,~\cite{iwasaki1982Raman, cingolani1987raman, roubi1988resonance, ibanez2018high, neal2021Raman, peng2019, peng2021, Grzeszczyk2022} the peaks that appear at 260.8~cm$^{-1}$ and 337.4~cm$^{-1}$ can be attributed correspondingly to the in-plane E$_{\textrm{g}}$ and out-of-plane A$_{\textrm{1g}}$ modes.
The peak at 134.4~cm$^{-1}$ is referred in this work to as $\omega_{\textrm{1}}$.
As its assignment in the literature is not well established (see Ref.~\cite{neal2021Raman, Grzeszczyk2022}), the possible origin is discussed in the following.
In addition, two separate peaks, denoted A$_{\textrm{{2u}}}\textrm{(TO)}$ and A$_{\textrm{{2u}}}\textrm{(LO)}$, can be distinguished on the low energy side of the A$_{\textrm{1g}}$ mode with Raman shifts equal 303.3~cm$^{-1}$ and 321.0~cm$^{-1}$.
Although the latter peak was associated with A$_\textrm{2u}$ phonon modes, the former one has not been reported so far. 
Their attribution is further analyzed in the text.

\begin{figure}[t]
    \centering
    \includegraphics[width=1\linewidth]{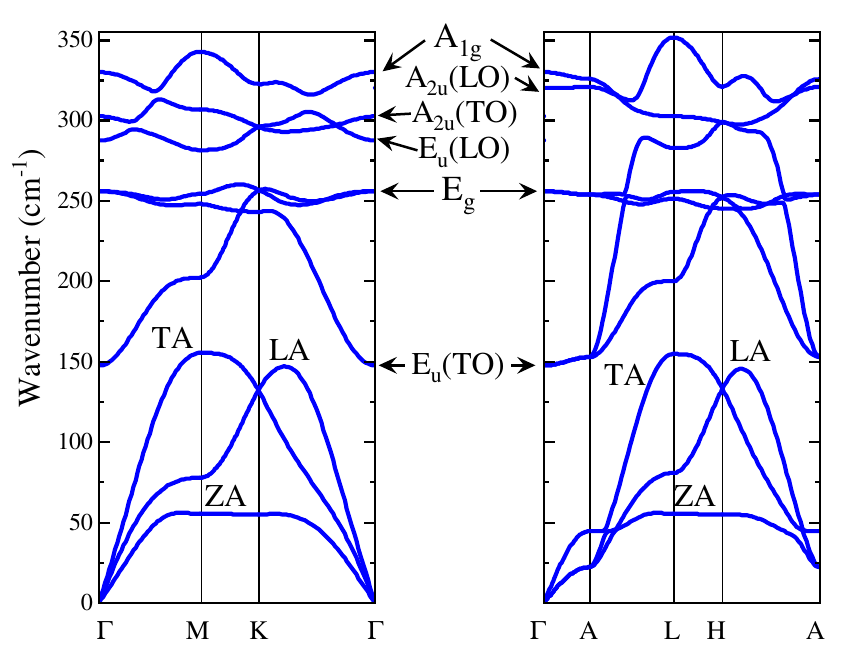}
    \caption
    {\label{fig:Phonons_disp} 
    Calculated phonon dispersion curves for bulk \HfS.
    The dispersion is split into two parts: $\Gamma$-M-K-$\Gamma$ and $\Gamma$-A-L-H-A, for clarity reason.}
\end{figure}

To support our investigation of Raman spectra measured in \HfS, we performed DFT calculations of the dispersion of phonons in this material.
The obtained result is presented in Fig.~\ref{fig:Phonons_disp}.
There are three branches of acoustic vibrations: longitudinal (LA), transverse (TA), and out-of-plane (ZA) in \HfS.
Several optical modes can be appreciated at Raman shifts higher than 150~cm$^{-1}$.
The Raman-active A$_\textrm{1g}$ and E$_\textrm{g}$ modes are characterized by wavenumbers of 330.0~cm$^{-1}$ and 256.0~cm$^{-1}$ at the $\Gamma$ point of the Brillouin zone (BZ), respectively.
The calculated values are in good agreement with the aforementioned experimental ones. 
The characteristic LO-TO splitting of the IR-active E$_u$ (E$_{\textrm{{u}}}\textrm{(LO)}$ and E$_{\textrm{{u}}}\textrm{(TO)}$) and A$_{2u}$ (A$_{\textrm{{2u}}}\textrm{(LO)}$ and A$_{\textrm{{2u}}}\textrm{(TO)}$) modes can also be appreciated.
The calculated wavenumbers at the $\Gamma$ point of the BZ equal to: E$_{\textrm{{u}}}\textrm{(TO)}$ -- 147.6~cm$^{-1}$, E$_{\textrm{{u}}}\textrm{(LO)}$ -- 287.5~cm$^{-1}$, A$_{\textrm{{2u}}}\textrm{(TO)}$  -- 302.5~cm$^{-1}$, and A$_{\textrm{{2u}}}\textrm{(LO)}$  -- 320.0~cm$^{-1}$.

%%%%%%%%%%%% {Temperature-dependent Raman scattering} %%%%%%%%%%%
\subsection{Temperature-dependent Raman scattering}

\begin{figure}[t]
    \centering
    \includegraphics[width=1\linewidth]{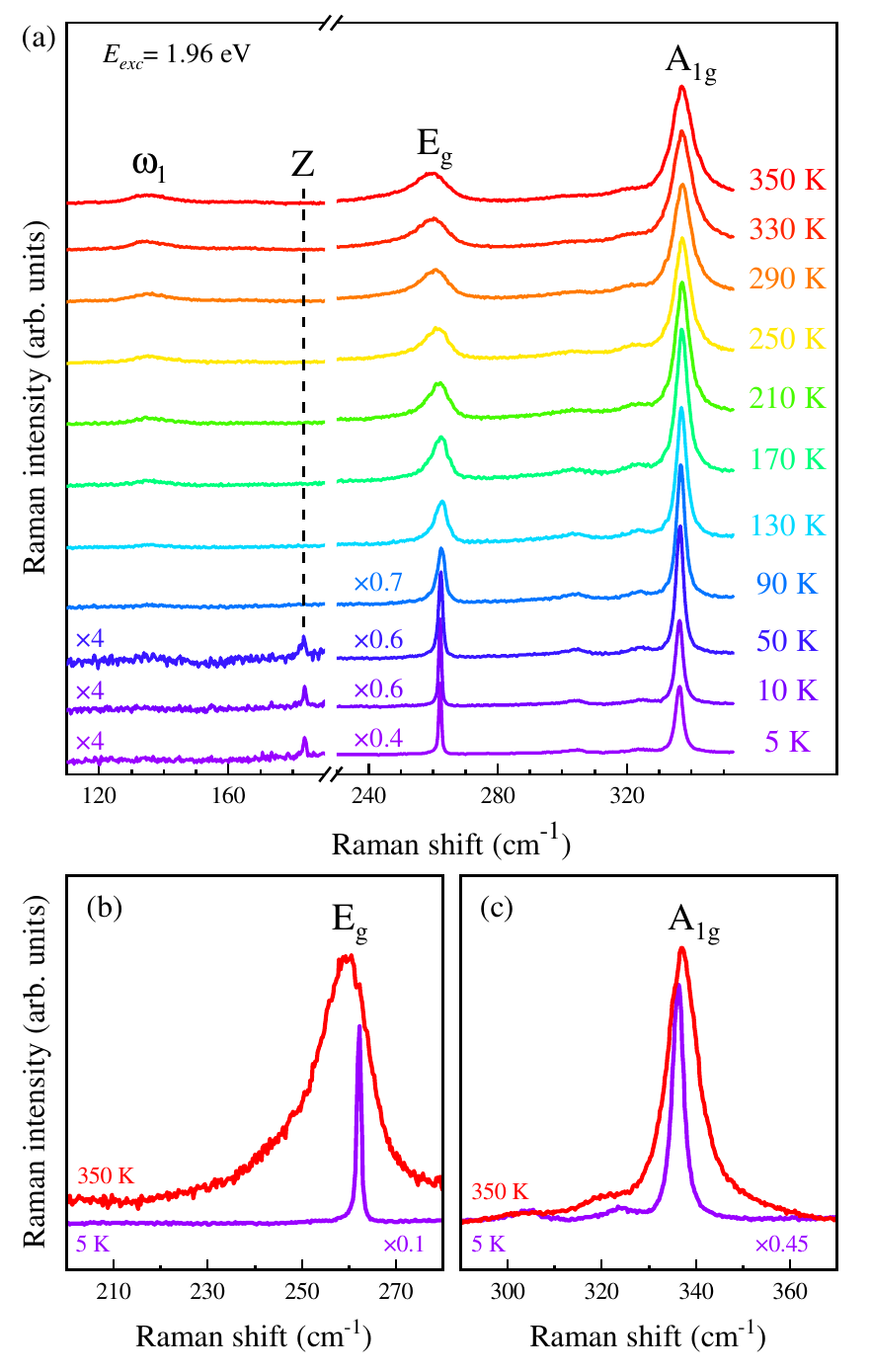}
    \caption{\label{fig:Raman_temp}
    (a) The temperature evolution of the unpolarized RS spectra measured on \HfS bulk under excitation of $E_{exc}$=1.96 eV.
    Note that the spectra were multiplied by scaling factors in order to better visualize some part of the RS spectra at given temperatures.
    The scaling factor, indicated in the Figure, correspond to the specific parts of the spectra before and after break of the x axis.
    For temperatures lower than 100 K a new mode, called here $Z$, has been observed and the quenching of $\omega_1$.
    Raman spectra of the (b) E$_{\textrm{g}}$ and (c) A$_{\textrm{1g}}$ modes measured at $T$=5~K and 350~K. 
    The x-axis ranges are the same in both the (b) and (c) panels, while the y-axis range is adjusted to maximize the intensities of the E$_{\textrm{g}}$ and A$_{\textrm{1g}}$ modes at $T$=350~K.
    The 5~K spectra are multiplied by the denoted numbers for clarity.}
\end{figure}

The evolution of the RS spectra as a function of temperature from $T$=5~K to $T$=350 K is shown in Fig.~\ref{fig:Raman_temp}(a). 
As can be seen in the Figure, the temperature dependence of the E$_{\textrm{g}}$ and A$_{\textrm{1g}}$ peaks can be followed for the whole temperature range.
Moreover, the low-energy satellites of the A$_{\textrm{1g}}$, labeled A$_{\textrm{{2u}}}\textrm{(TO)}$ and A$_{\textrm{{2u}}}\textrm{(LO)}$ peaks in Fig.~\ref{fig:Raman_300K}(b), are also apparent for the whole temperature range.
Interestingly, the shape of the low-frequency range of the RS spectra, $i.e.$ about 110~cm$^{-1}$ -- 190~cm$^{-1}$, is much more affected by the increase of the sample temperature. 
The narrow peak, denoted Z, is seen at $T$=5~K.
This peak vanishes quickly with increasing temperature, which is accompanied by the appearance of the $\omega_{\textrm{1}}$ mode.
The former peak is observed up to $T$=350~K.
In the first view, the most conspicuous effect of temperature on the RS spectra is the significant broadening of the E$_{\textrm{g}}$ and A$_{\textrm{1g}}$ peaks with increasing temperature due to the anharmonic effects.
In Fig.~\ref{fig:Raman_temp}(b) and (c), we compare RS spectra showing these two peaks, measured at $T$=5~K and $T$=350~K.
To describe the influence of temperature on the linewidths of the E$_{\textrm{g}}$ and A$_{\textrm{1g}}$ peaks, we fitted them with the Lorentz functions.
The full width at half maximum (FWHM) of the E$_{\textrm{g}}$ peak increases of around 16 times, from $\sim$0.9~cm$^{-1}$ at 5~K to $\sim$14.7~cm$^{-1}$ at 350~K.
The corresponding change in A$_{\textrm{1g}}$ FWHM is only approximately 3 times, from $\sim$2.7~cm$^{-1}$ at 5~K to 8.2~cm$^{-1}$ at 350~K.
Note that around the M point of the BZ (Fig.~\ref{fig:Phonons_disp}), there are two very flat branches, ZA and E$_\textrm{u}$(TO), at 50~cm$^{-1}$ and 200 cm~cm$^{-1}$, respectively. 
So, there could be a highly efficient channel of phonon scattering for the E$_{\textrm{g}}$ mode ($\sim$250~cm$^{-1}$), which could decay into pairs of ZA and E$_\textrm{u}$(TO) phonons. 
This is not so evident for the A$_{\textrm{1g}}$ mode, and this is probably why it has a longer lifetime, $i.e.$ a larger linewidth.
Moreover, it should be noted that the E$_{\textrm{g}}$ shape at $T$=350~K is asymmetric, which may suggest the appearance of additional RS peaks on its low-energy side due to the increased population of phonons at higher temperatures.

\begin{figure}[t]
    \centering
    \includegraphics[width=1\linewidth]{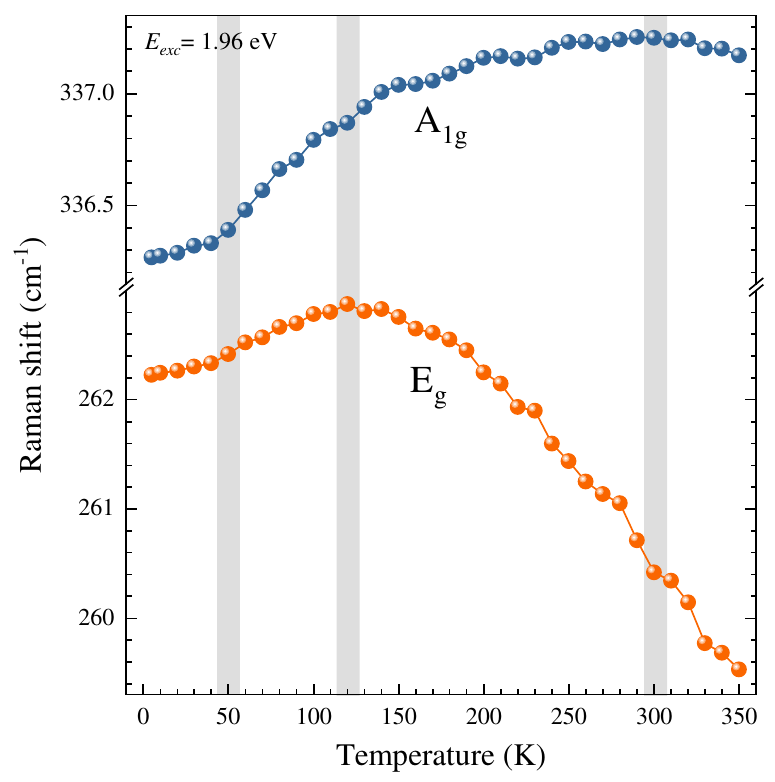}
    \caption{\label{fig:Ag_Eg_temp} 
    Temperature evolution of the Raman shift of the A$_{\textrm{1g}}$ and E$_{\textrm{g}}$ modes under excitation of $E_{exc}$=1.96~eV.
    The grey shaded areas indicate the "transition" temperatures.}
\end{figure}

In order to investigate the observation in details, the temperature evolution of the Raman shift of the four studied Raman peaks, $i.e.$ E$_{\textrm{g}}$, A$_{\textrm{1g}}$, $\omega_1$, and Z, we fitted them using Lorentz functions in the whole temperature range.
Fig.~\ref{fig:Ag_Eg_temp} presents the obtained temperature evolution of the Raman shifts of the A$_{\textrm{1g}}$ and E$_{\textrm{g}}$ modes.
As can be appreciated from the Figure, their temperature dependences are nontrivial. 
The A$_{\textrm{1g}}$ peak experiences a significant blueshift, of about 1~cm$^{-1}$, in the almost the whole temperature range.
Its initial increase is almost linear up to around $T$=50~K, further the shift transforms into a kind of logarithmic dependence with the maximum Raman shift at around $T$=300~K, and finally a small red shift is observed at the highest temperatures. 
The evolution of the E$_{\textrm{g}}$ mode is nonmonotonic.
The initial blue shift of the E$_{\textrm{g}}$ peak up to about $T$=50~K is also linear, as for the A$_{\textrm{1g}}$ mode, then the blue shift slope changes with a kind of maximum value at around $T$=120~K.
At higher temperatures, the typical red shift of the E$_{\textrm{g}}$ peak with increasing temperature is observed.

\begin{figure}[b]
    \centering
    \includegraphics[width=1\linewidth]{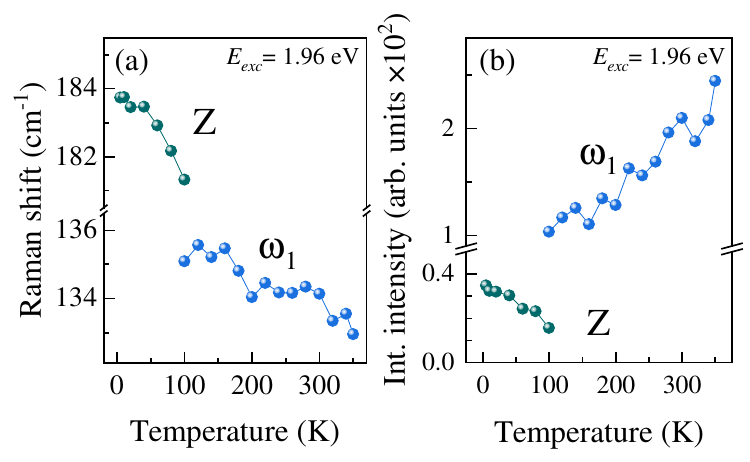}
    \caption{\label{fig:Z_omega}
    Temperature evolution of the (a) Raman shift and (b) integrated intensity of the $\omega_1$ and $Z$ modes.}
\end{figure}

To determine the assignment of the $\omega_1$ and $Z$ peaks, we show their Raman shifts and integrated intensities in Figs.~\ref{fig:Z_omega}(a) and (b), respectively.
The temperature dependence of the $\omega_1$ and Z frequencies is typical, both of them show red shifts for the entire temperature range.
The integrated intensity of the Z peak decreases with increasing temperature.
For temperatures higher than $T$=100~K, the Z mode is not apparent, while the $\omega_1$ peak emerges and its intensity grows up to $T$=350~K.

%%%%%%%%%%%%%%% {Polarization-resolved Raman scattering} %%%%%%

\subsection{Polarization-resolved Raman scattering}

\begin{figure}[t]
    \centering
    \includegraphics[width=1\linewidth]{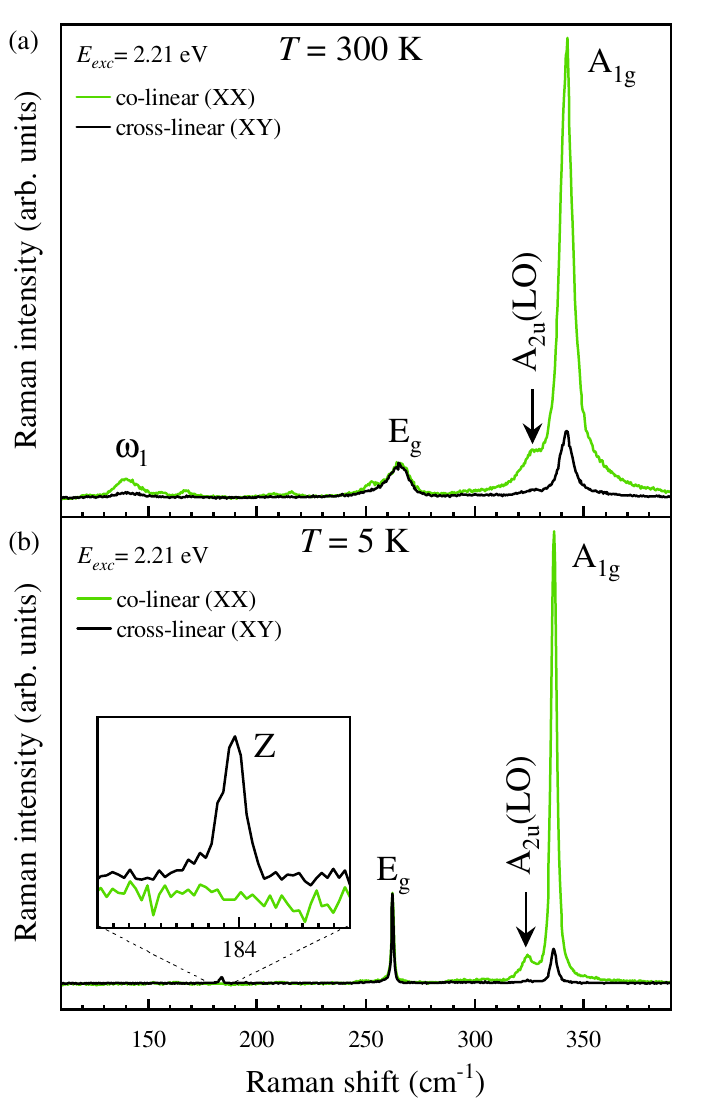}
    \caption{\label{fig:Polarization} 
    Polarization-resolved RS spectra of \HfS bulk for co- and cross-linear configurations measured at (a) $T$=300~K and (b) $T$=5~K. 
    Measurements were carried out under excitation of $E_{exc}$=2.21~eV.}
\end{figure}

In order to analyze polarization dependences of phonon modes, let us introduce the Raman tensors in our experimental backscattering arrangement.
First, we note that the vectors indicating the direction of the incoming and scattered light are along the $c$ direction of the crystal (perpendicular to the layers planes).
We perform our measurements in two configurations, co- and cross-linear, which correspond to the parallel (XX) and perpendicular (XY) orientations of the polarization of the incoming and scattered light. 
The Raman scattering intensity of a Raman-active mode in a crystal can be expressed by the Raman tensor ($R$) (see Ref.~\cite{Zhang2015} for details). 
For Raman-active  A$_{\textrm{1g}}$ and E$_\textrm{g}$ modes in 1T \HfS, their $R$ are listed as follows:~\cite{Zhang2015}
\begin{equation*}
\textrm{A}_{\textrm{1g}}: \left(\begin{array}{ccc}
 a & 0 & 0 \\
0 & a & 0 \\
0 & 0 & b 
\end{array}\right); \textrm{E}_{\textrm{g}}: \left(\begin{array}{ccc}
c & 0 & 0 \\
0 & -c & d \\
0 & d & 0 
\end{array}\right),
\left(\begin{array}{ccc}
0 & -c & -d \\
-c & 0 & 0 \\
-d & 0 & 0 
\end{array}\right).
\end{equation*}
Consequently, A$_{\textrm{1g}}$ mode is observed in the XX configuration and is absent in the XY configuration, while the E$_\textrm{g}$ mode is present in both the XX and XY configurations.
The results of the polarization-resolved RS measurements in the co- and cross-linear arrangements at $T$=300 K and $T$=5 K are shown in Figs.~\ref{fig:Polarization}(a) and (b), respectively.
It can be seen that the A$_{\textrm{1g}}$ mode is strongly attenuated in the cross-linear configuration, but does not completely quench. 
The similar behavior can be observed for the A$_{\textrm{{2u}}}\textrm{(LO)}$ and $\omega_1$ modes.
On the contrary, the E$_{\textrm{g}}$ peak has the same intensity for both co- and cross-linear polarizations.
Finally, the surprising behavior of the Z mode observed at $T$=5~K can be noticed, which can be observed only for cross-linear configuration. 
Note that the comprehensive analysis of the orientation of the excitation polarization in relation to the crystallographic orientation of the investigated \HfS is presented in the SI.

%%%%%%%%%%%%%%% {Excitation energy-dependent Raman scattering} %%%%%%
\subsection{Excitation energy-dependent Raman scattering}

\begin{figure*}[ht]
    \centering
    \includegraphics[width=1\linewidth]{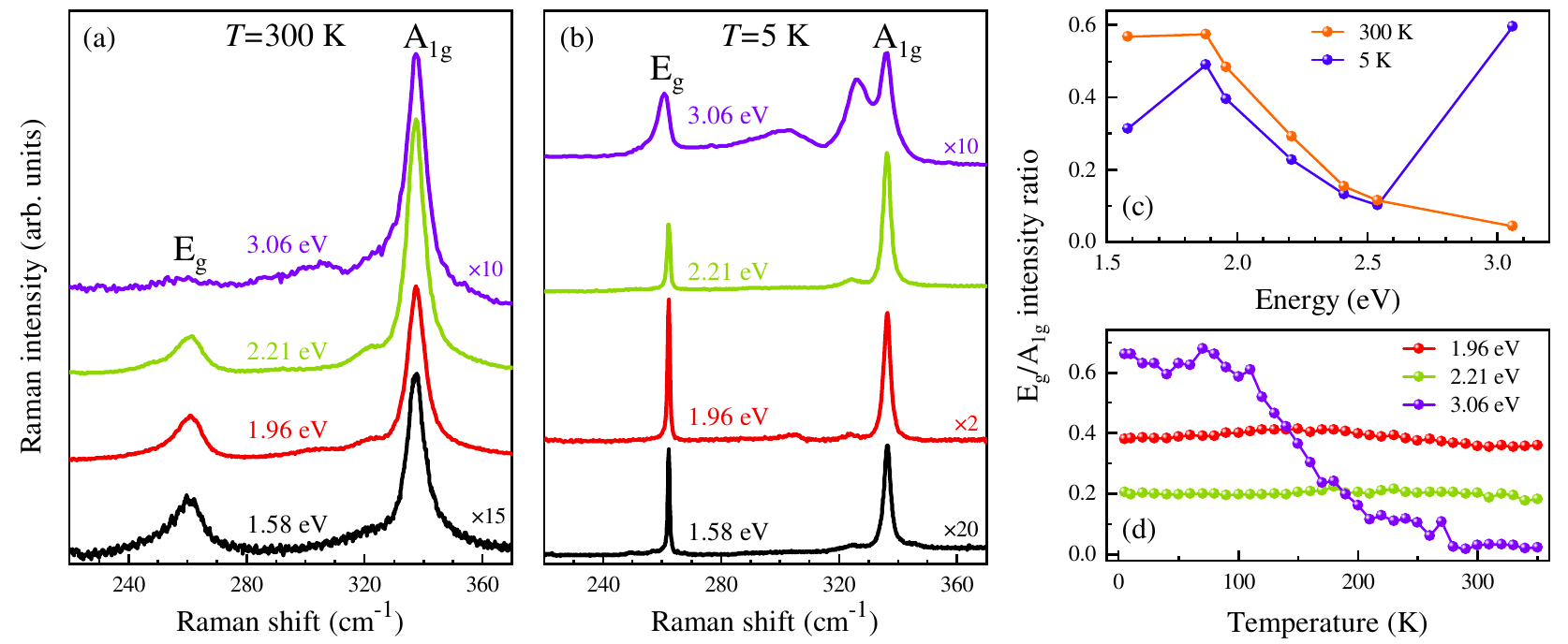}
  	\caption{\label{fig:Excitation_energy}
	The Raman scattering spectra measured on Hf$_2$ bulk at (a) $T$=300~K and (b) $T$=5~K under a series of different excitation energies, indicated in the Figure. 
	The spectra are multiplied by scaling factors for clarity.
	The integrated relative E$_\textrm{g}$/A$_{\textrm{1g}}$ intensity as a function of: (c) excitation energy, measured at $T$=5~K and $T$=300~K and (d) temperature, measured with excitation of 1.96~eV, 2.21~eV, and 3.06 eV.}
\end{figure*}

The RS is an inelastic light scattering process, which, in principle, involves a virtual intermediate state.
More complicated is the description of the process in which the excited state coincides with a non-negligible density of electronic states in a process referred to as the resonant RS (RRS).
The light-matter interaction in semiconductors measured by RRS can be modulated by the variation of the excitation energy and/or temperature.~\cite{Carvalho2015, Soubelet2016, Molas2017, golasa2017resonance, Osiekowicz2021, Bhatnagar2022} 
The RS spectra excited with a series of lasers and measured at $T$=300~K and $T$=5~K are presented in Figs.~\ref{fig:Excitation_energy}(a) and (b), respectively. 
It can be appreciated from the Figure that both the temperature at which the experiment was performed as well as the excitation energy of the laser have a substantial effect on the shape of the RS spectra.
First, we address the evolution of the Raman spectra measured at room temperature.
The RS signal is more than an order of magnitude larger under excitations of 1.96~eV and 2.21~eV than the corresponding ones obtained with excitations of 1.58~eV and 3.06~eV.
Furthermore, the shape of the RS spectrum excited with 3.06~eV is distinct, with the apparent quenching of the E$_{\textrm{g}}$ mode.
The analogous analysis of the results acquired at $T$=5~K brings us to the conclusion that the change of temperature affects not only the integrated intensity of the Raman signal but also the relative E$_\textrm{g}$/A$_{\textrm{1g}}$ intensity (compare the RS spectra measured under 1.96~eV at $T$=300~K and 5~K).
For all excitations, except $E_{exc}$=3.06~eV,  the RS spectra are dominated by relatively narrow E$_{\textrm{g}}$ and A$_{\textrm{1g}}$ modes.
The RS spectrum to $T$=5 K under the 3.06~eV excitation is much richer, as two additional features are observed close to the A$_{\textrm{1g}}$ mode at 302.6~cm$^{-1}$ and 326.2~cm$^{-1}$. 
To investigate the observed effect in detail, we plot the relative intensity E$_\textrm{g}$/A$_{\textrm{1g}}$ ratio for both temperature (5~K and 300~K) as a function of the used excitation energies, see Figs.~\ref{fig:Excitation_energy}(c).
At room temperature, the initial constant E$_{\textrm{g}}$/A$_{\textrm{1g}}$ ratio on the level of about 0.6 under excitations of 1.58~eV and 1.88~eV is followed by its gradual decrease to almost 0 under excitation of 3.06~eV.
On the contrary, the corresponding evolution obtained at $T$=5~K is almost the same, except for the case when the RS signal is excited with 3.06~eV.
The influence of temperature on the E$_{\textrm{g}}$/A$_{\textrm{1g}}$ ratio is presented in Fig.~\ref{fig:Excitation_energy} for three excitation energies: 1.96~eV, 2.21~eV, and 3.06~eV.
It is seen that the E$_{\textrm{g}}$/A$_{\textrm{1g}}$ ratio is almost constant over the whole temperature range from $T$=5~K to 350~K, while there is a significant variation under the 3.06~eV excitation.
At the lowest temperature up to around 100~K, the ratio is fixed to a level of about 0.6.
The following increase in temperature leads to a gradual decrease of the E$_{\textrm{g}}$/A$_{\textrm{1g}}$ ratio, which finally reaches almost zero for temperatures above approx. 280~K.

%%%%%%%%%%%%%%% {Discussion} %%%%%%
\section{Discussion \label{sec:Disscusion}}

First, we will analyze the properties of two main vibrational modes, $i.e.$ E$_{\textrm{g}}$ and A$_{\textrm{1g}}$, which are Raman active.
As can be seen in Fig.~\ref{fig:Ag_Eg_temp}, the most interesting characteristic of the E$_{\textrm{g}}$ and A$_{\textrm{1g}}$ modes is their temperature evolution, which is nonmonotonic.
Typically, the temperature increase leads to the monotonic red shift of phonon energies originated by the expected spring constant reduction due to lattice expansion, as observed in classical semiconductor, $i.e.$ silicon,~\cite{Balkanski1983} or layered vdW materials, $e.g.$ MoS$_2$, WS$_2$, black phosphorus, and SnSe.~\cite{Pawbake2016, Lapinska2016, Buruiana2022}
The observed temperature evolution of the E$_{\textrm{g}}$ and A$_{\textrm{1g}}$ peak energies in \HfS is more complex.
Three "transition" temperatures can be distinguished at which the evolution changes its slope and/or also sign, which are of about 50~K, 120~K, and 300~K.
Note that similar temperatures have been reported in the literature.~\cite{neal2021Raman, peng2021}
The transition at the highest temperature of about 300~K was ascribed to the reversible phase transition that occurs in \HfS between the 1T polytype present below $T$=300~K and 3R-T polytype apparent above $T$=300~K.~\cite{peng2021}
For the 1T phase, the Hf and S atoms of the neighboring layers are placed on top of each other, forming AAA stacking, while there is a linear shift of the consecutive layers in the 3R-T phase, resulting in ABC stacking.
The A$_{\textrm{1g}}$ blue shift in the 1T phase was associated with a disproportionate thermal expansion behavior between intra- and inter-layer thicknesses in the multi-trilayer structure.~\cite{peng2021}
The origin of the "transition" temperatures of 50~K and 120~K is more intriguing.
S.~N.~Neal et al.~\cite{neal2021Raman} found two "transition" temperatures in the Raman spectra of \HfS, which were attributed to local lattice distortions involving a slight motion of the S centers with respect to the Hf ions so as to change the bond lengths and angles a little while maintaining the same overall space group.
On the other hand, we found that iodine (I$_2$) molecules are located in the vdW gaps between the layers of the studied \HfS crystals (see Ref.~\citenum{Zawadzka2022} for details).
Consequently, we propose that the non-trivial temperature evolution of the E$_{\textrm{g}}$ and A$_{\textrm{1g}}$ peaks can be described in terms of local lattice distortion due to the presence of the I$_2$ molecules.
Note that more sophisticated theoretical calculations are required to investigate this scenario, which, however, are beyond the scope of this experimental study.
As seen in Fig.~\ref{fig:Polarization}, the E$_{\textrm{g}}$ peak is observed in both polarization configurations, which supports its correspondence to the aforementioned Raman tensor.
For the A$_{\textrm{1g}}$ peak, a non-negligible intensity detected in the cross-linear configuration is observed.
This may suggest that the Hf and S atoms are not arranged in the perfect 1T phase, even at $T$=5~K (compare Figs.~\ref{fig:Polarization}(a) and (b)), due to the presence of iodine molecules.

Let us discuss the origin of two spectral features observed at lower side of the A$_\textrm{1g}$ mode, which are referred to as A$_{\textrm{{2u}}}\textrm{(TO)}$ and A$_{\textrm{{2u}}}\textrm{(LO)}$.
Because of the symmetry, these modes are infrared-active and should not be active in the RS experiment.  
The presence of IR-active modes in the Raman spectra can be related to strain, disorder, or Coulomb interactions.\cite{jae2017}
In fact, the IR-active modes were apparent in the RS spectra in several TMDs due to the presence of symmetry breaking, disorder, or resonant conditions of RS excitation.\cite{golasa2015disorder,  Nam2015, lee2015, Molas2017, ibanez2018high, Bhatnagar2022}. 
It should be noted that some works attributed the peak at 321 cm$^{-1}$ to the E$_\textrm{u}(\textrm{LO})$ mode.~\cite{neal2021Raman, iwasaki1982Raman, cingolani1987raman}.
As it was shown theoretically and experimentally in Ref.~~\citenum{ibanez2018high}, the Grüneisen and pressure coefficients of the A$_{\textrm{{2u}}}\textrm{(LO)}$ mode differ from those calculated to E$_\textrm{u}(\textrm{LO})$.
This supports our attribution of the peak appearing at 321.0~cm$^{-1}$ to the A$_{\textrm{{2u}}}\textrm{(LO)}$.
The peak observed at 303.3~cm$^{-1}$ has not been reported so far.
Consequently, we assign this peak to the A$_{\textrm{{2u}}}\textrm{(TO)}$ mode using the phonon dispersion, shown in Fig.~\ref{fig:Phonons_disp}.

The last open question is the possible assignment of the $\omega_1$ and Z peaks observed correspondingly in the higher and lower ranges of temperatures, see Fig.~\ref{fig:Raman_temp}(a). 
In the case of the $\omega_1$ peak, a long dispute about its origin can be found in the literature.~\cite{roubi1988resonance, ibanez2018high, peng2019, neal2021Raman, Grzeszczyk2022}
Due to our previous results of the RS experiment in hydrostatic pressure~\cite{Grzeszczyk2022} and the calculated phonon dispersion (see Fig.~\ref{fig:Phonons_disp}), the $\omega_1$ peak should be related to a second-order process.
There are several possibilities, such as E$_\textrm{u}(\textrm{LO})-\textrm{E}_\textrm{u}(\textrm{TO})$, A$_\textrm{1g}-\textrm{E}_\textrm{u}(\textrm{TO})(\textrm{M})$, and E$_\textrm{u}(\textrm{LO})(\textrm{M})-\textrm{LA(M)}$, whose theoretical positions are 140 cm$^{-1}$, 128 cm$^{-1}$, and 133 cm$^{-1}$, respectively.
Following the assignment given in Ref.~\citenum{peng2019}, we attribute the $\omega_1$ peaks to the differential combination of the E$_\textrm{u}(\textrm{LO})-\textrm{E}_\textrm{u}(\textrm{TO})$ phonons.
Although this attribution is not conclusive, the apparent quenching of the $\omega_1$ mode at low temperature, seen in Fig.~\ref{fig:Raman_temp}(a), supports its attribution to the difference process.
The origin of the Z peak is the most intriguing question to be addressed.
The Z peak is observed from $T$=5~K up to about $T$=100~K and is not present in the spectra excited with energies equal to 2.41 eV and 3.06 eV.
Moreover, the Z peak can only be observed in the cross-linear configuration (see Fig.~\ref{fig:Polarization}(b)), which is a typical behavior for modes of $B$-type symmetry.
These types of modes can be observed in materials that crystallize in $e.g.$ a distorted orthorhombic structure (D$^1$6$_{2h}$), $e.g.$ GeS, SnSe.~\cite{Zawadzka2021, Buruiana2022}
Due to the appearance of the Z peak at 5~K, we excluded the differential combination of phonons. 
Moreover, its linewidth at 5 K ($\sim1.1$~cm$^{-1}$) is comparable to the corresponding one of the E$_\textrm{g}$ peak ($\sim0.9$~cm$^{-1}$).
Knowing that the $E$ and $B$ modes are characterized by the in-plane vibrations of atoms without and with a selected direction, respectively, we tentatively assign the Z peak to the flat dispersion of the E$_\textrm{u}(\textrm{TO})$ mode at the M point of the BZ, see Fig.~\ref{fig:Phonons_disp}.
In our opinion, this type of vibration may be affected by the presence of I$_2$ molecules between the \HfS layers, which in-plane arrangement can be regular with a favored direction.
However, more experimental and theoretical studies are needed to confirm this attribution.

Finally, let us address the effect of resonant excitation on the measured RS spectra, shown in Fig.~\ref{fig:Excitation_energy}.
The resonant excitation may lead to a significant enhancement of the RS intensity in TMD as well as the activation of otherwise inactive modes.~\cite{grzeszczyk2016Raman, Bhatnagar2022} 
This offers supplementary information on the coupling of particular phonons to electronic transitions of a specific symmetry.~\cite{Carvalho2015}
The crossover between the non-resonant and resonant conditions can be achieved not only by the variation of the excitation energy but also by the modulation of temperature.~\cite{golasa2017resonance, Osiekowicz2021}
As is seen in Figs.~\ref{fig:Excitation_energy}(a) and (b) the RS signal is enhanced significantly for the 1.96~eV and 2.21~eV.
This leads to the conclusion that resonant conditions of RS are achieved under these excitations for \HfS due to the energy proximity of kind of direct absorption process near to the indirect band gap.~\cite{GREENAWAY1965, TERASHIMA1987, Gaiser2004, Zelewski2017}
The corresponding transitions, can be associated with the so-called "band nesting", which gives rise to singularity features in the joint density of states and is characterized by the strong light-matter interaction (see Refs.~\cite{Carvalho2013, Kozawa2014} for details).
We focus our analysis of the influence of the excitation energy on the relative E$_{\textrm{g}}$/A$_{\textrm{1g}}$ intensity, as it is completely free from other matters, $e.g.$ sensitivity of the experimental setup.
The variation of the excitation energy in the range from 1.58~eV to 2.54~eV affects the relative E$_{\textrm{g}}$/A$_{\textrm{1g}}$ ratio, but it almost stays at the same level for the measurements done at $T$=300~K and $T$=5~K, see Fig.~\ref{fig:Excitation_energy}(c).
This suggests that the resonant conditions of the RS excitation are similar for both the E$_{\textrm{g}}$ and A$_{\textrm{1g}}$ peaks.
However, there is a strong variation between the low and room temperatures for the 3.05~eV excitation.
The results, shown in Fig.~\ref{fig:Excitation_energy}(c), lead us to the following conclusions: (i) the excitation conditions of the RS spectra with the 1.96~eV and 2.21~eV do not depend on the temperature. 
It may suggest that a slight change in the electronic structure of \HfS as a function of temperature does not affect the overall excitation conditions; 
(ii) the resonant excitation conditions in the vicinity of 3.05 eV are composed of transitions of different symmetry. 
The E$_{\textrm{g}}$ and A$_{\textrm{1g}}$ modes correspond to the in-plane and out-of-plane types of sulfur vibrations, respectively. 
The enhancement of the  E$_{\textrm{g}}$/A$_{\textrm{1g}}$ ratio may suggest that the symmetry of the electronic bands involved in the resonance at low temperatures is characterized by the type of in-plane symmetry.
The resonance conditions due to these bands decrease significantly from about 100~K to 280~K.
At the highest temperature (>280~K), their influence on the enhancement of the E$_{\textrm{g}}$ peak is lost, and therefore only the A$_{\textrm{1g}}$ peak is apparent in the corresponding RS spectra.
The temperature evolution of the E$_{\textrm{g}}$/A$_{\textrm{1g}}$ ratio for the 3.06~eV excitation can be described in terms of quantum interference, as was reported for thin layers of MoTe$_2$,~\cite{miranda2017quantum} $i.e.$ the contributions to the Raman susceptibility from different regions (individual k-points) of the BZ add with particular signs (plus or minus). 
This may lead to enhancement or quenching of phonon modes, particularly with high-energy excitation, such as observed for 3.06 eV, compared to the band gap of \HfS due to the large number of possible direct excitations.

% {Conclusions}
\section{Conclusions \label{sec:Conclusions}}

Raman scattering in bulk \HfS is investigated as a function of temperature (5~K -- 350~K) with optical excitation of several excitation energies. 
An unexpected temperature dependence of the energies of main Raman-active (A$_{\textrm{1g}}$ and E$_{\textrm{g}}$) modes with the temperature-induced blueshift in the low-temperature limit is reported, which is beyond a simple model considering the thermal expansion of material.
The apparent quenching of the A$_{\textrm{1g}}$ mode at $T$=5~K and of the E$_{\textrm{g}}$ mode at $T$=300~K in the RS spectrum excited with 3.06~eV excitation is also observed, strongly suggesting the resonant character of electron-phonon interactions for that energy.
The low-temperature quenching of a mode $\omega_1$ (134 cm$^{-1}$) and the emergence of a new mode at approx. 184 cm$^{-1}$, labeled $Z$ are also reported.
The optical anisotropy of the RS in \HfS is also reported, which is highly susceptible to the excitation energy.
We discuss the results in the context of possible resonant character of light-phonon interactions.
Analyzed is also a possible effect of the iodine molecules intercalated in the van der Waals gaps between neighboring \HfS layers, which inevitably result from the growth procedure.
The presented results strongly point out the need for a more strict theoretical analysis, which is beyond the scope of this experimental report.

%{Methods}
\section*{Methods\label{sec:Methods}}
Single crystals of \HfS were synthesized in a two-zone chemical vapor transport (CVT) furnace. 
First, the precursor materials were sealed in quartz tubes 25 cm long and of diameter of 8 cm. 
Iodine ($\text{I}_{2}$ ) was used as a transport agent. 
The reaction and growth temperatures were set to 1050 K and 950 K, respectively.
The growth was carried out continuously for 120 hours. 
With a higher stability of \HfS against iodine, CVT growth starts first at elevated temperatures. 
In the vapor phase, HfI and HfI$_2$ can also dominate the direction of reaction. 
After 120 hours, the reaction was stopped automatically and the reactor was cooled down to room temperature in 5 hours. 
High-quality \HfS single crystals of 0.8-1 cm$^2$ size were grown in the low-temperature part of the reactor. 
The pattern of powder X-Ray diffraction (XRD) of the investigated material was found to be in perfect agreement with that of the octahedral 1T phase of \HfS.\cite{lukovsky1973IR}
A very minor amount of an unidentified phase was also detected in the crystal by the XRD technique.\cite{ibanezPrivate}

RS measurements were performed with excitation light focused by means of a 50x long-working distance objective with a 0.55 numerical aperture (NA) producing a spot of about 1~$\mu$m diameter. 
The signal was collected via the same microscope objective (backscattering geometry), sent through a 0.75~m monochromator, and then detected using a liquid nitrogen-cooled charge-coupled device (CCD) camera. 
Temperature-dependent experiments in the range from $T$=5~K to 350 K were done by placing the sample on a cold finger in a continuous flow cryostat mounted on $x$–$y$ motorized positioners. 
Excitation energy ($E_{exc}$)-dependent experiments were carried out using seven lasers with energy (wavelength): 1.58~eV (785~nm), 1.88~eV (660~nm), 1.96~eV (633~nm), 2.21~eV (561~nm), 2.41~eV (515~nm), 2.54~eV (488~nm), and 3.06~eV (405~nm).
The excitation power focused on the sample was kept at approx. 500~$\mu$W in all measurements.
Polarization-sensitive RS spectroscopy was realized in both co- (XX) and cross-linear (XY) configurations, corresponding to the parallel and perpendicular orientations of the excitation and detection polarization axes, respectively. 
To investigate the crystallographic anizotropy of the RS signal, a half-wave plate was mounted in both the excitation and detection paths to simultaneously rotate the polarization of the incoming light and outgoing signal.
Schematic illustration of the corresponding setup is presented in Supplementary Information (SI).

Density functional theory (DFT) calculations were conducted in Vienna Ab initio Simulation Package~\cite{VASP} with Projector Augmented Wave method.~\cite{PAW} 
Perdew–Burke–Ernzerhof parametrization~\cite{PBE} of general gradients approximation to the exchange-correlation functional was used. 
The plane waves basis cutoff energy was set to 550 eV and a 9$\times$9$\times$6 $\Gamma$-centered Monkhorst-Pack k-grid sampling was applied. Geometrical structure was optimized with $10^{-5}$ eV/\AA and 0.01 kbar criteria for the interatomic forces and stress tensor components, respectively. Grimme's D3 correction was applied to describe the interlayer vdW interactions.~\cite{D3} 
Spin-orbit interaction was taken into account during geometry optimization. Phonon calculations were performed within Parliński-Li-Kawazoe method,~\cite{Parlinski} as implemented in Phonopy software.~\cite{Phonopy} 
The 3$\times$3$\times$2 supercells were found sufficient to converge the interatomic force constants within the harmonic approximation.

\section*{Acknowledgment}
The work has been supported by the National Science Centre, Poland (grants no. 2017/27/B/ST3/00205 and 2018/31/B/ST3/02111). 
Z.M. and W.Z. acknowledge support from the National Natural Science Foundation of China (grant no. 62150410438), the International Collaboration Project (no. B16001), and the Beihang Hefei Innovation Research Institute (project no. BHKX-19-02).
DFT calculations were performed with the support of the PLGrid infrastructure.

\bibliographystyle{apsrev4-2}
\bibliography{biblio}

\newpage
\onecolumngrid
\setcounter{figure}{0}
\setcounter{section}{0}
\renewcommand{\thefigure}{S\arabic{figure}}
\renewcommand{\thesection}{S\Roman{section}}
\include{si_arxiv}

\end{document}

%% file: si_arxiv.tex
	\begin{center}
	%%%%%%%%% ABSTRACT TITLE
	{\large{{\bf  \textsc{Supplementary Information}} \\ The effect of temperature and excitation energy on Raman scattering in bulk \HfS}}
	%%%%%%%%% ABSTRACT AUTHORS
\end{center}

%{Angular intensity}
\section{Polarization sensitive measurements of Raman scattering}

\begin{figure}[h]
    \centering
    \includegraphics[width=.4\linewidth]{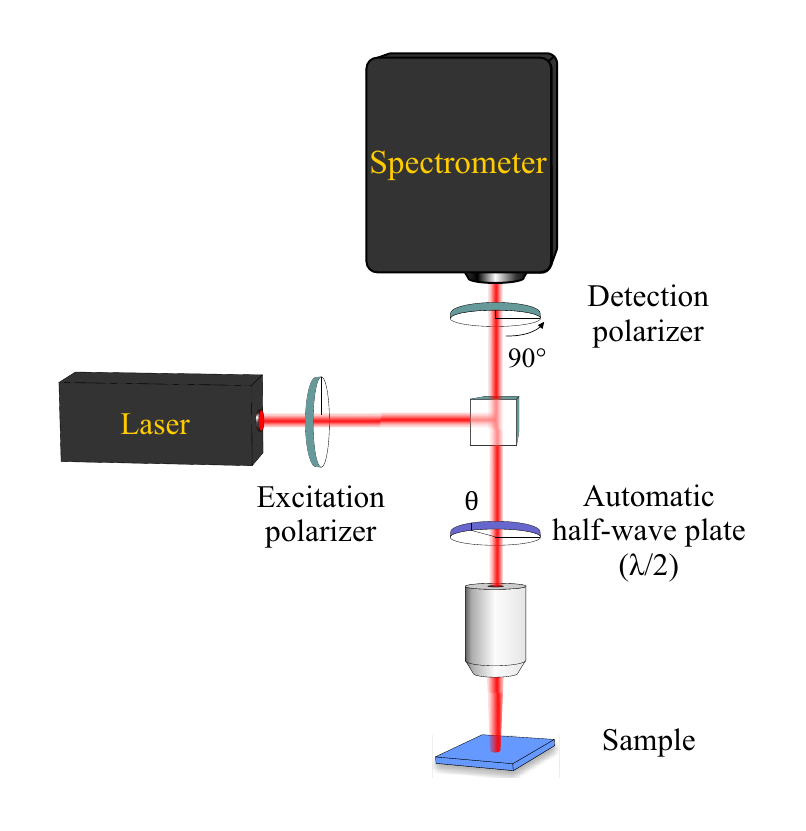}
    \caption{\label{fig:s1} 
    Schematic illustration of the experimental setup used for the measurements of polarization-resolved measurements with reference to crystal orientation.
    The polarization axis of the excitation polarizer is fixed, whereas the polarization axis of the detection polarized can be aligned in the same direction or in perpendicular.
    The half-wave plate is motorized.}
\end{figure}

Schematic illustration of the experimental setup used for the measurements of polarization-resolved Raman scattering (RS) with reference to crystal orientation is presented in Fig.~\ref{fig:s1}. 
The optical setup consists of two polarizers and a half-wave plate. 
One polarizer is placed in the excitation path, directly after the laser output, while the second one is placed in the detection path, just before the spectrometer. 
The relative orientation of polarization axes of the excitation and detection polarizers permits to choose between co- and cross-linear configurations of experiments. 
The co-linear configuration is reached when both polarizers are aligned in the same direction, whereas for the cross-linear configuration, their polarization axes are perpendicular to each other. 
The rotation of a half-wave plate, mounted in the joint part of the excitation and detection paths, allows to rotate simultaneously both the excitation and detection polarizations in reference to the studied sample.
This type of experiment is an analog of the rotation of the sample, and it has the advantage that the rotation axis of the sample does not need to be centered on the excitation/detection spot.
The used half-wave plate is motorized and controlled by a computer, which allows automatic measurements of polarization-resolved spectra.
Note that the aforementioned arrangement of the experimental setup allows us to probe the in-plane (ani)isotropy of the sample, while polarization-dependent results, presented in the main text (Fig.~6), were recorded without the half-wave plate.

\section{Polarization-resolved Raman spectra}

%Through polarization-sensitive RS we studied the anisotropy of \HfS presented by A$_{\textrm{1g}}$ and E$_\textrm{g}$. In fig.~\ref{}.a is shown the spectra for co- and cross-linear configuration at 300 K excited by 2.21 eV (500 $\mu$W). $A_{1g}$ mode was hard attenuated in cross-linear configuration, but not complete extinct. The same behavior can be observed for $A_{2u}(LO)$ (321 cm$^{-1}$) supporting its assignment as a mode with $A$ instead $E$. Following the same trend $\omega_1$ is attenuated for this configuration too. Interestingly, together $E_g$ band, in a co-linear configuration is possible to observe a shoulder, close to 250 cm$^{-1}$ suggesting an $A$ symmetry mode in this region. For 5 K experiment (fig.~\ref{fig:6}.b) we can observe almost the same situation for $A_{1g}$ and $E_g$ ($\omega_1$ is not present for 5 K), but surprising $Z$ mode is present only for cross-linear configuration, reveling an anti-symmetry like $B$ modes.

\begin{figure*}[b]
    \centering
    \includegraphics[width=1\linewidth]{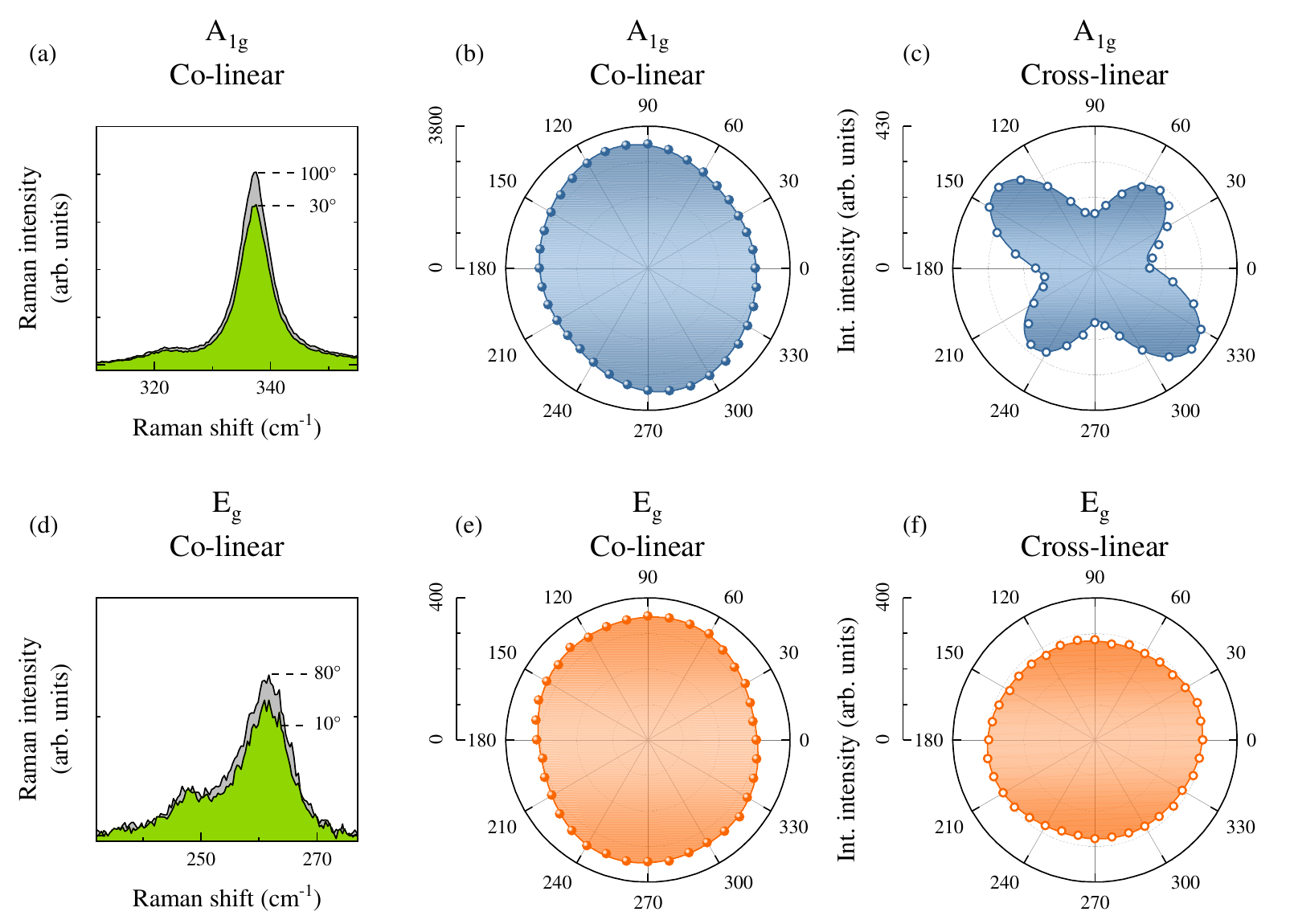}
    \caption{\label{fig:anis250K} 
    Co-polarized RS spectra of the (a) A$_{\textrm{1g}}$ and (d) E$_\textrm{g}$ peaks recorded for two relative angles between the excitation polarization and the \HfS orientation.
    Polar plots of the integrated intensities of the phonon modes: (b), (c) A$_{\textrm{1g}}$ and (e), (f) E$_\textrm{g}$, measured on \HfS in the (b), (e) co- and (c), (f) cross-linear configurations at $T$=250 K under $E_{exc.}$=2.21~eV.
    The solid curves correspond to the fits using Eq.~\ref{eq1}.}
\end{figure*}

In order to study the in-plane anisotropy of \HfS, measurements of the polarization resolved RS spectra were performed as a function of an angle between the excitation polarization and the crystallographic orientation of the \HfS crystal. 
Figs.~\ref{fig:anis250K}(a) and (d) present, respectively, the co-linear polarized RS spectra of the A$_{\textrm{1g}}$ and E$_\textrm{g}$ peaks measured for angles, which correspond to the maximum and minimum intensities. 
For isotropic layered materials, $e.g.$ graphene and MoS$_2$,~\cite{Lee2014, Kim2020} there is no dependence of the intensity of the $A$ and $E$ modes for the co-linear configuration with the rotation of the crystal along the axis perpendicular to the layers planes.
In contrast, for anisotropic layered materials, $e.g.$ black phosphorus, GeS, and SnSe,~\cite{Kim2020, Zawadzka2021, Xu2021, Buruiana2022} there is a strong dependence of the intensities of the phonon modes on the crystallographic orientation of the studied material.
As can be appreciated in Figs.~\ref{fig:anis250K}(a) and (d), there is a small but non-negligible difference in the intensity of both the A$_{\textrm{1g}}$ and E$_\textrm{g}$ peaks measured for the selected values of rotation angle.
Figs.~\ref{fig:anis250K}(b)-(c) and (e)-(f) present polar plots of the integrated intensities as a function of the detection angle for the A$_{\textrm{1g}}$ and E$_\textrm{g}$ modes in the co- and cross-linear configurations, respectively. 
The solid curves, seen in the Figure, represent fits of the modes intensities as a function of light polarization, $I(\theta)$, described by:~\cite{Ribeiro2015}
\vspace{-15pt}
\begin{center}
\begin{equation}\label{eq1}
     I(\theta)= (|A|sin^2(\theta-\phi)+|C|cos(\alpha)cos^2(\theta-\phi))^2 + |C|^2sin^2(\alpha) cos^4(\theta-\phi)  
\end{equation}
\end{center}
\noindent where $|A|$ and $|C|$ are the amplitudes of the phonon modes, $\phi$ represents the phase of polarization dependence, $\alpha$ represents the phase difference. 
As can be seen in Figs.~\ref{fig:anis250K}(b), (e), and (f), the presented polarization evolution is characterized a slight deformation of the "circular" behavior.
Moreover, in the case of the cross-linear evolution of the A$_{\textrm{1g}}$, the observed 4-fold symmetry is not isotropic, $i.e.$ the intensities of the perpendicular "arms" are different. 
These results suggest that the studied \HfS crystal at $T$=250~K deviate form a perfect 1T phase.
In our opinion, these small deformations can be attributed to the presence of the iodine (I$_2$) molecules, located in the vdW gaps between the layers of the studied \HfS crystals (see main text for details).

\begin{figure}[t]
    \centering
    \includegraphics[width=1\linewidth]{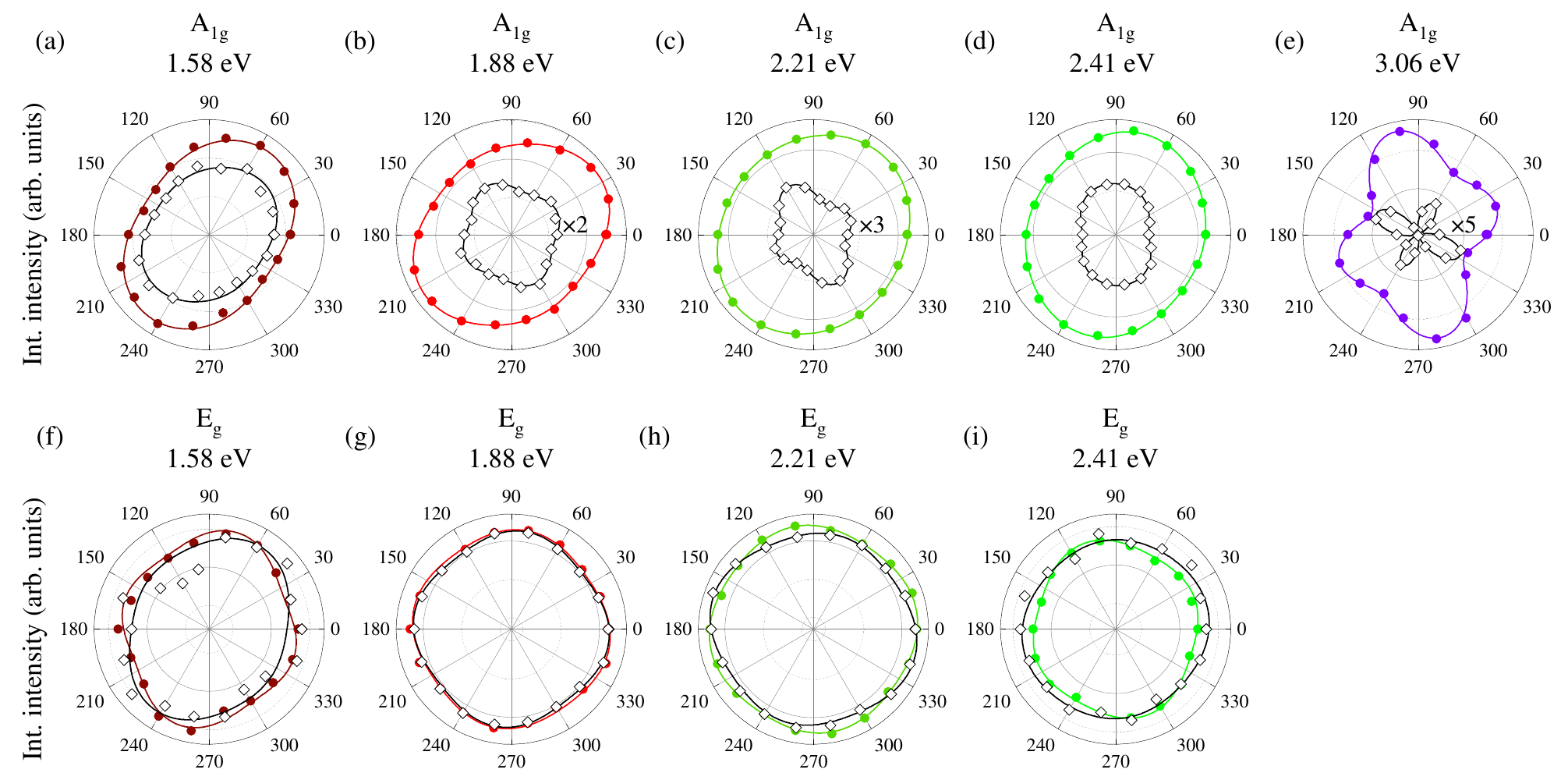}
    \caption{\label{fig:excanis} 
     Polar plots of the integrated intensities of the phonon modes: (a)-(e) A$_{\textrm{1g}}$ and (f)-(i) E$_\textrm{g}$, where co- (full points) and cross-linear (open points) configurations were measured on HfS$_2$ at $T$=300 K under a series of excitation energies.
    The solid curves correspond to the fits using Eq.~\ref{eq1}.}
\end{figure}

In order to examine the effect of the excitation energy on the polarization properties of phonon modes, we performed the polarization-resolved RS experiments under five different excitations (1.58 eV, 1.88 eV, 2.21 eV, 2.41 eV, and 3.06 eV) at room temperature. 
Fig.~\ref{fig:excanis} presents the corresponding polarization-sensitive RS polar plots obtained for the A$_{\textrm{1g}}$ and E$_\textrm{g}$ modes with a series of excitation energies for co- (full) points and cross-linear (open points) configurations.
In the co-linear configuration, the A$_{\textrm{1g}}$ mode displays 2-fold symmetry with an angle period of 180$^\circ$ for almost all excitation energies, except for the 3.06~eV.
Surprisingly, there is a 4-fold symmetry of the A$_{\textrm{1g}}$ mode for a co-circular configuration with the 3.06~eV excitation.
The analogous situation occurs for the corresponding cross-circular configuration of the A$_{\textrm{1g}}$ mode.
Although 4-fold symmetry is expected for cross-linear configuration, seen in panels (b), (c), and (e) of Fig.~\ref{fig:excanis}, there is also a 2-fold symmetry, apparent for the 1.58~eV and 2.41~eV excitations.
Moreover, not only the shape of the polarization evolution of the A$_{\textrm{1g}}$ mode varies, but also the orientation of the polarization axes.
In contrast, the effect of the excitation energies on the polarization dependence of the E$_\textrm{g}$ mode is much weaker, see Figs.~\ref{fig:excanis}(f)-(i).
In all the shown cases, the observed shape is almost "circular".
Note that the observed inﬂuence of the excitation energies on the axes and shape of the polarization properties of the phonon modes is very similar to those reported for different anisotropic layered materials, $e.g.$ black phosphorus, GeS, and SnSe.~\cite{Kim2015, Zawadzka2021, Buruiana2022}
To conclude, except for 3.06 eV, the excitation energy does not significantly affect the RS shape in \HfS, see Fig.~7 in the main text, but it can modify the polarization properties of phonon modes with reference to the orientation of the sample.
This suggests that the electron-phonon coupling in the studied \HfS~is quite strong.
Moreover, as it depends on mode symmetries, compare results for the A$_{\textrm{1g}}$ and E$_\textrm{g}$ modes presented in Fig.~\ref{fig:excanis}, the symmetries of the involved electronic bands are essential in the electron-phonon interactions and may affect the polarization properties of modes.